\documentclass[runningheads]{llncs}

\usepackage{stmaryrd,amsmath,amssymb,newlfont,graphicx,verbatim}
\usepackage[ruled,lined,boxed,commentsnumbered,linesnumbered]{algorithm2e}
\usepackage{subfigure}
\usepackage{fancyvrb}

\newcommand{\flow}{\mathsf{flow}}
\newcommand{\jump}{\mathsf{jump}}
\newcommand{\inv}{\mathsf{inv}}
\newcommand{\init}{\mathsf{init}}

\newcommand{\reach}{\mathsf{Reach}}
\newcommand{\goal}{\mathsf{goal}}

\newcommand{\lrf}{\mathcal{L}_{\mathbb{R}_{\mathcal{F}}}}
\newcommand{\citep}{\cite}

\newcommand{\hide}[1]{}
\newcommand{\eg}{{\em e.g.}}
\newcommand{\ie}{{\em i.e.}}
\newcommand{\enforce}{\mathsf{enforce}}

\title{Parameter Synthesis for Cardiac Cell Hybrid Models Using $\delta$-Decisions}

\titlerunning{Analyzing cardiac cell dynamics using $\delta$-decisions}

\authorrunning{Liu et al.}
\author{Bing Liu\inst{1}
\and Soonho Kong\inst{1}
\and Sicun Gao\inst{1}
\and Paolo Zuliani\inst{2}
\and Edmund M. Clarke\inst{1}}
\institute{Computer Science Department, Carnegie Mellon University, USA
\and School of Computing Science, Newcastle University, UK}

\begin{document}

\maketitle
\begin{abstract} 
A central problem in systems biology is to identify parameter values such that a biological model satisfies some behavioral constraints (\eg, time series). In this paper we focus on parameter synthesis for hybrid (continuous/discrete) models, as many biological systems can possess multiple operational modes with specific continuous dynamics in each mode. These biological systems are naturally modeled as hybrid automata, most often with nonlinear continuous dynamics. However, hybrid automata are notoriously hard to analyze --- even simple reachability for hybrid systems with linear differential dynamics is an undecidable problem. In this paper we present a parameter synthesis framework based on $\delta$-complete decision procedures that sidesteps undecidability.
We demonstrate our method on two highly nonlinear hybrid models of the cardiac cell action potential. 
The results show that our parameter synthesis framework is convenient and efficient, and 
it enabled us to select a suitable model to study and identify crucial parameter ranges related to cardiac disorders.
\end{abstract}

\section{Introduction}

Computational modeling and analysis methods are playing a crucial role in understanding the complex dynamics of biological systems \citep{liu12jbcb}. In this paper we address the parameter synthesis problem for hybrid models of biological systems.
This problem amounts to finding sets of parameter values for which a model satisfies some precise 
behavioral constraints, such as time series or reachability properties. We focus on hybrid 
continuous/discrete models, since one of the key aspects of many biological systems is their differing 
behavior in various `discrete' modes. For example, it is well-known that the five stages of the cell 
cycle are driven by the activation of different signaling pathways. Hence, hybrid system models are often 
used in systems biology (see, \eg, 
\citep{chen04,tomlin04,Hu04,ye08,aihara10,antoniotti03,lincoln04,baldazzi11}).

Hybrid systems combine discrete control computation with continuous-time evolution. The state space 
of a hybrid system is defined by a finite set of continuous variables and modes. In each mode, the
continuous evolution ({\em flow}) of the system is usually given by the solution of ordinary differential
equations (ODEs). At any given time a hybrid system dwells in one of its modes and each variable 
evolves accordingly to the flow in the mode. Jump conditions control the switch to another mode,
possibly followed by a `reset' of the continuous variables. Thus, the temporal dynamics of a 
hybrid system is piecewise continuous.

Hybrid models of biological systems often involve many parameters such as rate constants of 
biochemical reactions, initial conditions, and threshold values in jump conditions. Generally, only 
a few rate constants will be available or can be measured experimentally --- in the latter case
the rate constants are obtained by fitting the model to experimental observations. Furthermore, 
it is also crucial to figure out what initial conditions or jump conditions may lead to an unsafe
state of the system, especially when studying hybrid systems used to inform clinical 
therapy \citep{tanaka10}. All such questions fall within the \textit{parameter synthesis} problem,
which is extremely difficult for hybrid systems. Even simple reachability questions for hybrid 
systems with linear (differential) dynamics are undecidable \citep{henzinger96}.
Therefore, the parameter synthesis problem needs to be relaxed in a sound manner in order to solve
it algorithmically --- this is the approach we shall follow.

In this paper, we tackle the parameter synthesis problem using $\delta$-complete procedures \cite{gao12a} 
for deciding first-order formula with arbitrary computable real functions, including solutions 
of Lipschitz-continuous ODEs \citep{gao12b}. Such procedures may return answers with
one-sided $\delta$-bounded errors, thereby overcoming undecidability issues (note that the maximum 
allowable error $\delta$ is an arbitrarily small positive rational). In our approach we describe 
the set of states of interest as a first-order logic formula and perform bounded model checking \cite{BMC}
to determine reachability of these states. We then adapt an interval constrains propagation based algorithm 
to explore the parameter space and identify the sets of resulting parameters. 
We show the applicability of our method by carrying out a thorough case study characterized by 
highly nonlinear hybrid models. We discriminate two cardiac cell action potential 
models \cite{fenton98,orovio08} in terms of cell-type specificity and identify parameter ranges for which a cardiac cell may lose excitability.
The results show that our method can obtain biological insights that are consistent with experimental observations, 
and scales to complex systems. 
In particular, the analysis we carried out in the cardiac case study is difficult to be performed by --- if not out of the scope of ---  state-of-the-art tools \cite{prism,breach,taliro,biocham}.

\paragraph{Related Work.}
A survey of modeling and analysis of biological systems using hybrid models can be found in \cite{luca08}. Analyzing the properties of biochemical networks using formal verification techniques is being actively pursued by a number of researchers, for which we refer to Brim's {\em et al.} recent survey \cite{BrimSFM13}. Of particular interest in our context are parameter synthesis methods for qualitative behavior specifications (\eg, temporal logic formulas). The method introduced in \cite{rovergene} can deal with parameter synthesis for piecewise affine linear systems. For ODEs, Donz\'{e} {\em et al.}~\cite{donze} explore the parameter space using adaptive sampling and simulation, while Palaniappan {\em et al.} \cite{liu13} use a statistical model checking approach. Other techniques perform a sweep of the entire (bounded) parameter space, after it has been discretized \cite{Calzone06,Donaldson08}. Randomized optimization techniques were used for parameter estimation in stochastic hybrid systems \cite{Koutroumpas08}, while identification techniques for affine systems were used in \cite{Cinquemani08}. The techniques above can handle nonlinear hybrid systems only through sampling and 
simulation, and so are incomplete. Our approach is instead $\delta$-complete. It is based on verified numerical integration and constraint programming algorithms, which effectively enable an over-approximation of the system dynamics to be computed. Thus, if a model is found to be unfeasible (i.e. an $\mathsf{unsat}$ answer is returned, see Section 2 for more details), then this is correct. This behavior better fits with the safety requirements expected by formal verification.


\section{$\delta$-Decisions for Hybrid Models}
We encode reachability problems of hybrid automata using a first-order language $\lrf$ over the reals, 
which allows the use of a wide range of real functions including nonlinear ODEs. 
We then use $\delta$-complete decision procedures to find solutions to these formulas to synthesize 
parameters. 

\begin{definition}[$\lrf$-Formulas]
Let $\mathcal{F}$ be a collection of computable real functions. We define:
\begin{align*}
t& := x \; | \; f(t(\vec x)), \mbox{ where }f\in \mathcal{F} \mbox{ (constants are 0-ary functions)};\\
\varphi& := t(\vec x)> 0 \; | \; t(\vec x)\geq 0 \; | \; \varphi\wedge\varphi
\; | \; \varphi\vee\varphi \; | \; \exists x_i\varphi \; |\; \forall x_i\varphi.
\end{align*}
\end{definition}
By computable real function we mean Type 2 computable, which informally requires that a (real) 
function can be algorithmically evaluated with arbitrary accuracy. Since in general 
$\lrf$ formulas are undecidable, the decision problem needs to be relaxed. In particular, for 
any $\lrf$ formula $\phi$ and any rational $\delta >0$ one can obtain a $\delta$-weakening 
formula $\phi^\delta$ from $\phi$ by substituting the atoms $t > 0$ with $t > -\delta$ (and
similarly for $t \geq 0$). Obviously, $\phi$ implies $\phi^\delta$, but not the {\em vice versa}.
Now, the $\delta$-decision problem is deciding correctly whether:
\begin{itemize}
	\item $\phi$ is false ($\mathsf{unsat}$);
	\item $\phi^\delta$ is true ($\delta$-$\mathsf{sat}$).
\end{itemize}
If both cases are true, then either decision is correct. In previous work~\cite{gao12a,gao12b,gao13}
we presented algorithms ($\delta$-{\em complete} decision procedures) for solving $\delta$-decision 
problems for $\lrf$ and for ODEs. These algorithms have been implemented in the dReal 
toolset \cite{dreal}. More details on $\delta$-decision problems are in Appendix. 

Now we state the encoding for hybrid models. Recall that hybrid automata generalize finite-state
automata by permitting continuous-time evolution (or {\em flow}) in each discrete state (or {\em mode}). 
Also, in each mode an {\em invariant} must be satisfied by the flow, and mode switches are controlled
by {\em jump} conditions.
\begin{definition}[$\lrf$-Representations of Hybrid Automata]\label{lrf-definition}
A hybrid automaton in $\lrf$-representation is a tuple
\begin{multline*}
H = \langle X, Q, \{{\flow}_q(\vec x, \vec y, t): q\in Q\},\{\inv_q(\vec x): q\in Q\},\\
\{\jump_{q\rightarrow q'}(\vec x, \vec y): q,q'\in Q\},\{\init_q(\vec x): q\in Q\}\rangle
\end{multline*}
where $X\subseteq \mathbb{R}^n$ for some $n\in \mathbb{N}$, $Q=\{q_1,...,q_m\}$ is a finite set of modes, and the other components are finite sets of quantifier-free $\lrf$-formulas.
\end{definition}
\begin{example}[Nonlinear Bouncing Ball]
The bouncing ball is a standard hybrid system model. It can be $\lrf$-represented in the following way:
\begin{itemize}
\item $X = \mathbb{R}^2$ and $Q = \{q_u, q_d\}$. We use $q_u$ to represent bounce-back mode and $q_d$ the falling mode.
\item $\flow = \{\flow_{q_u}(x_0, v_0, x_t, v_t, t), \flow_{q_d}(x_0, v_0, x_t, v_t, t)\}$. We use $x$ to denote the height of the ball and $v$ its velocity. Instead of using time derivatives, we can directly write the flows as integrals over time, using $\lrf$-formulas:
\begin{itemize}
\item $\flow_{q_u}(x_0, v_0, x_t, v_t, t)$ defines the dynamics in the bounce-back phase:
$$(x_t = x_0 + \int_0^{t} v(s) ds) \wedge (v_t = v_0 + \int_0^t g(1-\beta v(s)^2) ds)$$
\item $\flow_{q_d}(x_0, v_0, x_t, v_t, t)$ defines the dynamics in the falling phase:
$$(x_t = x_0 + \int_0^{t} v(s) ds) \wedge (v_t = v_0 + \int_0^t g(1+\beta v(s)^2) ds)$$
\end{itemize}where
$\beta$ is a constant. Again, note that the integration terms define Type 2 computable functions.
\item $\jump = \{\jump_{q_u \rightarrow q_d} (x, v, x', v'), \jump_{q_d \rightarrow q_u} (x, v, x', v')\}$ where
\begin{itemize}
\item $\jump_{q_u \rightarrow q_d} (x, v, x', v')$ is $(v= 0 \wedge x' = x \wedge v' = v)$.
\item $\jump_{q_d \rightarrow q_u} (x, v, x', v')$ is $(x= 0 \wedge v' = \alpha v\wedge x'=x)$,  for some constant $\alpha$.
\end{itemize}
\item $\init_{q_d}$ is $(x=10 \wedge v=0)$ and $\init_{q_u}$ is $\bot$.
\item $\inv_{q_d}$ is $(x>=0 \wedge v>=0)$ and $\inv_{q_u}$ is $(x>=0 \wedge v<=0)$.
\end{itemize}
\end{example}

We now show the encoding of bounded reachability, which is used for encoding the parameter synthesis
problem. We want to decide whether a given 
hybrid system reaches a particular region of its state space after following a (bounded) number
of discrete transitions, \ie, jumps. First, we need to define auxiliary formulas used
for ensuring that a particular mode is picked at a certain step.
\begin{definition}
Let $Q = \{q_1,...,q_m\}$ be a set of modes. For any $q\in Q$, and $i\in\mathbb{N}$, use  $b_{q}^i$ to represent a Boolean variable. We now define
$$\enforce_Q(q,i) = b^i_{q} \wedge \bigwedge_{p\in Q\setminus\{q\}}\neg b^{i}_{p}$$
$$\enforce_Q(q, q',i) = b^{i}_{q}\wedge \neg b^{i+1}_{q'} \wedge \bigwedge_{p\in Q\setminus\{q\}} \neg b^i_{p} \wedge \bigwedge_{p'\in Q\setminus\{q'\}} \neg b^{i+1}_{p'}$$
We omit the subscript $Q$ when the context is clear.\end{definition}

We can now define the following formula that checks whether a {\em goal} region of the automaton
state space is reachable after exactly $k$ discrete transitions. We first state 
the simpler case of a hybrid system without invariants.
\begin{definition}[$k$-Step Reachability, Invariant-Free Case]
Suppose $H$ is an invariant-free hybrid automaton, $U$ a subset of its state space represented by $\goal$,
and $M>0$. The formula $\reach_{H,U}(k,M)$ is defined as:
\begin{eqnarray*}
& &\exists^X \vec x_{0} \exists^X\vec x_{0}^t\cdots \exists^X \vec x_{k}\exists^X\vec x_{k}^t\exists^{[0,M]}t_0\cdots \exists^{[0,M]}t_k.\\
& &\bigvee_{q\in Q} \Big(\init_{q}(\vec x_{0})\wedge \flow_{q}(\vec x_{0}, \vec x_{0}^t, t_0)\wedge \enforce(q,0)\Big)\\
\wedge & & \bigwedge_{i=0}^{k-1}\bigg( \bigvee_{q, q'\in Q} \Big(\jump_{q\rightarrow q'}(\vec x_{i}^t, \vec x_{i+1})\wedge \enforce(q,q',i)\\
& & \hspace{4.5cm}\wedge\flow_{q'}(\vec x_{i+1}, \vec x_{i+1}^t, t_{i+1})\wedge \enforce(q',i+1)\Big)\bigg)\\
\wedge & &\bigvee_{q\in Q} (\goal_q(\vec x_{k}^t)\wedge \enforce(q,k))
\end{eqnarray*}
where $\exists^X x$ is a shorthand for $\exists x\in X$.
\end{definition}
Intuitively, the trajectories start with some initial state satisfying $\init_q(\vec x_{0})$ for some $q$. 
Then, in each step the trajectory follows $\flow_q(\vec x_{i}, \vec x_{i}^t, t)$ and makes a continuous flow from $\vec x_i$ to $\vec x_i^t$ after time $t$. When the automaton makes a $\jump$ from mode $q'$ to $q$, it resets variables following $\jump_{q'\rightarrow q}(\vec x_{k}^t, \vec x_{k+1})$. The auxiliary $\enforce$ formulas ensure that picking $\jump_{q\rightarrow q'}$ in the $i$-the step enforces picking $\flow_q'$ in the $(i+1)$-th step.

When the invariants are not trivial, we need to ensure that for all the time points along a continuous flow, the invariant condition holds. We need to universally quantify over time, and the encoding is as follows:
\begin{definition}[$k$-Step Reachability, Nontrivial Invariant]\label{br2}
Suppose $H$ contains invariants, and $U$ is a subset of the state space represented by $\goal$. The $\lrf$-formula $\reach_{H,U}(k,M)$ is defined as:
\begin{eqnarray*}
& &\exists^X \vec x_{0} \exists^X\vec x_{0}^t\cdots \exists^X \vec x_{k}\exists^X\vec x_{k}^t \exists^{[0,M]}t_0\cdots \exists^{[0,M]}t_k.\\
& &\bigvee_{q\in Q} \Big(\init_{q}(\vec x_{0})\wedge \flow_{q}(\vec x_{0}, \vec x_{0}^t, t_0)\wedge \enforce(q,0)\\
& &\hspace{5cm} \wedge \forall^{[0,t_0]}t\forall^X\vec x\;(\flow_{q}(\vec x_{0}, \vec x, t)\rightarrow \inv_{q}(\vec x))\Big) \\
\wedge & &\bigwedge_{i=0}^{k-1}\bigg( \bigvee_{q, q'\in Q} \Big(\jump_{q\rightarrow q'}(\vec
x_{i}^t, \vec x_{i+1})\wedge \flow_{q'}(\vec x_{i+1}, \vec x_{i+1}^t, t_{i+1})\wedge \enforce(q,q',i)\\
& & \hspace{1.5cm}\wedge\enforce(q',i+1)\wedge \forall^{[0,t_{i+1}]}t\forall^X\vec x\;(\flow_{q'}(\vec x_{i+1}, \vec x,
t)\rightarrow \inv_{q'}(\vec x)) )\Big)\bigg)\\
\wedge & &\bigvee_{q\in Q} (\goal_q(\vec x_{k}^t)\wedge \enforce(q,k)).
\end{eqnarray*}
\end{definition}
The extra universal quantifier for each continuous flow expresses the requirement that for all the time points between the initial and ending time point ($t\in[0,t_i+1]$) in a flow, the continuous variables $\vec x$ must take values that satisfy the invariant conditions $\inv_q(\vec x)$.

\paragraph{Parameter Identification.}
The parameter identification problem we tackle is basically a $k$-step reachability question: Is there a parameter
combination for which the model reaches the goal region in $k$ steps? If none exists, then the model is 
{\em unfeasible}. Otherwise, a witness (\ie, a value for each parameter) is returned. Note that because we ask for
$\delta$-decisions, the returned witness might correspond to a {\em spurious} behavior of the system. The occurrence
of such behaviors can be controlled via the precision $\delta$, but in general cannot be eliminated. 
We have developed the dReach tool (\verb#http://dreal.cs.cmu.edu/dreach.html#) 
that automatically builds reachability formulas from a hybrid model and a goal 
description. Such formulas are then verified by our $\delta$-complete solver dReal \citep{dreal}.

\section{Case Study}

To exemplify different aspects of our parameter synthesis framework, we carried out a case study on
models of cardiac cell electrical dynamics. All experiments reported below were done
using a machine with an Intel Core i5 3.4GHz processor and 8GB RAM. The precision $\delta$
was set to $10^{-4}$. The model files are available at \verb#http://www.cs.cmu.edu/~liubing/cmsb14/#.

\subsection{Hybrid models of cardiac cells}
The heart rhythm is enabled by the electrical activity of cardiac muscle cells, which make up the atria and ventricles. The electrical dynamics of cardiac cells is governed by the organized opening and closing of ion channel gates on the cell membrane. Improper functioning of the cardiac cell ionic channels can cause the cells to lose excitability, which disorders electric wave propagation and leads to cardiac abnormalities such as ventricular \textit{tachycardia} or \textit{fibrillation}. In order to understand the mechanisms of cardiac disorders,
hybrid automata models have been recently developed, including the Fenton-Karma (FK) model \cite{fenton98} and the Bueno-Cherry-Fenton (BCF) model \cite{orovio08}.
\begin{figure}[t]
\centering
\includegraphics[scale=0.52]{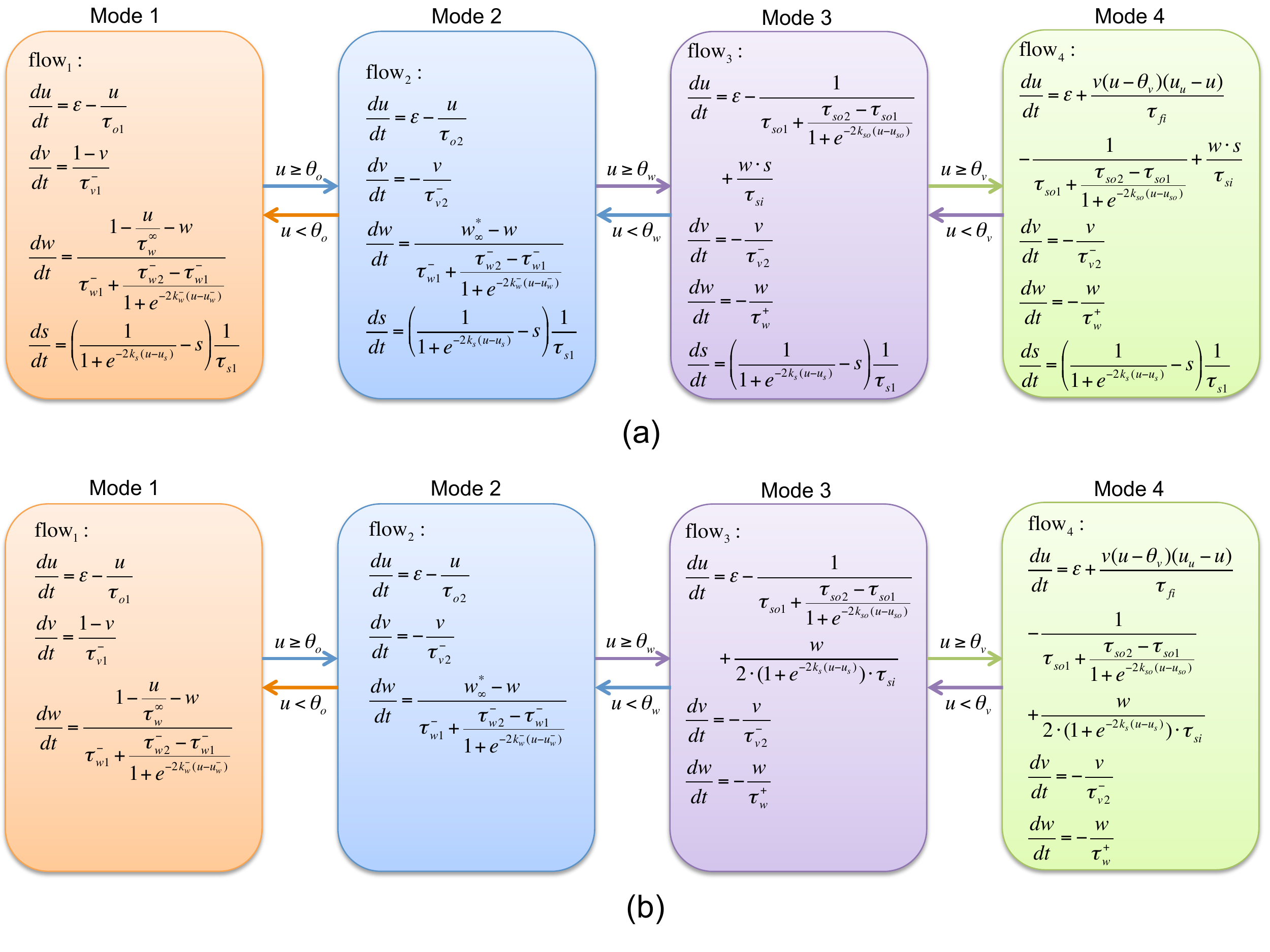}
\caption{Hybrid models of cardiac cells. (a) BCF model. (b) FK model.}
\label{model}
\end{figure}
\paragraph{BCF Model.}
In this model, the change of cells transmembrane potential $u$, in response to an external stimulus $\epsilon$ from neighboring cells, is regulated by a fast ion channel gate $v$ and two slow gates $w$ and $s$.
Figure \ref{model}(a) shows the four modes associated with the BCF model. In Mode $1$, gates $v$ and $w$ are open and gate $s$ is closed. The transmembrane potassium current causes the decay of $u$. The cell is resting and waiting for stimulation. We assume an external stimulus $\epsilon$ equal to $1$ that lasts for $1$ millisecond. The stimulation causes $u$ to increase, which may trigger $\jump_{1 \rightarrow 2}: u \geq \theta_o$. When this jump takes place, the system switches to Mode 2 and $v$ starts closing, and the decay rate of $u$ changes. The system will jump to Mode 3 if $u \geq \theta_w$. In Mode 3, $w$ is also closing; $u$ is governed by the potassium current and the calcium current. When $u \geq \theta_v$, Mode 4 can be reached, which signals a successful action potential (AP) initiation. In Mode 4, $u$ reaches its peak due to the fast opening of the sodium channel. The cardiac muscle contracts and $u$ starts decreasing.
\paragraph{FK Model.}
As shown in Figure \ref{model}(b), this model comprises the same four modes and equations of the BCF model, except that the current change induced by gate $s$ is reduced to an explicit term which is integrated in the right-hand side of $du/dt$. Similarly to the BCF model, an AP can be successfully initiated when Mode 4 is reached.

\begin{figure}[thb]
\centering
\subfigure[]{
  \includegraphics[width=8cm]{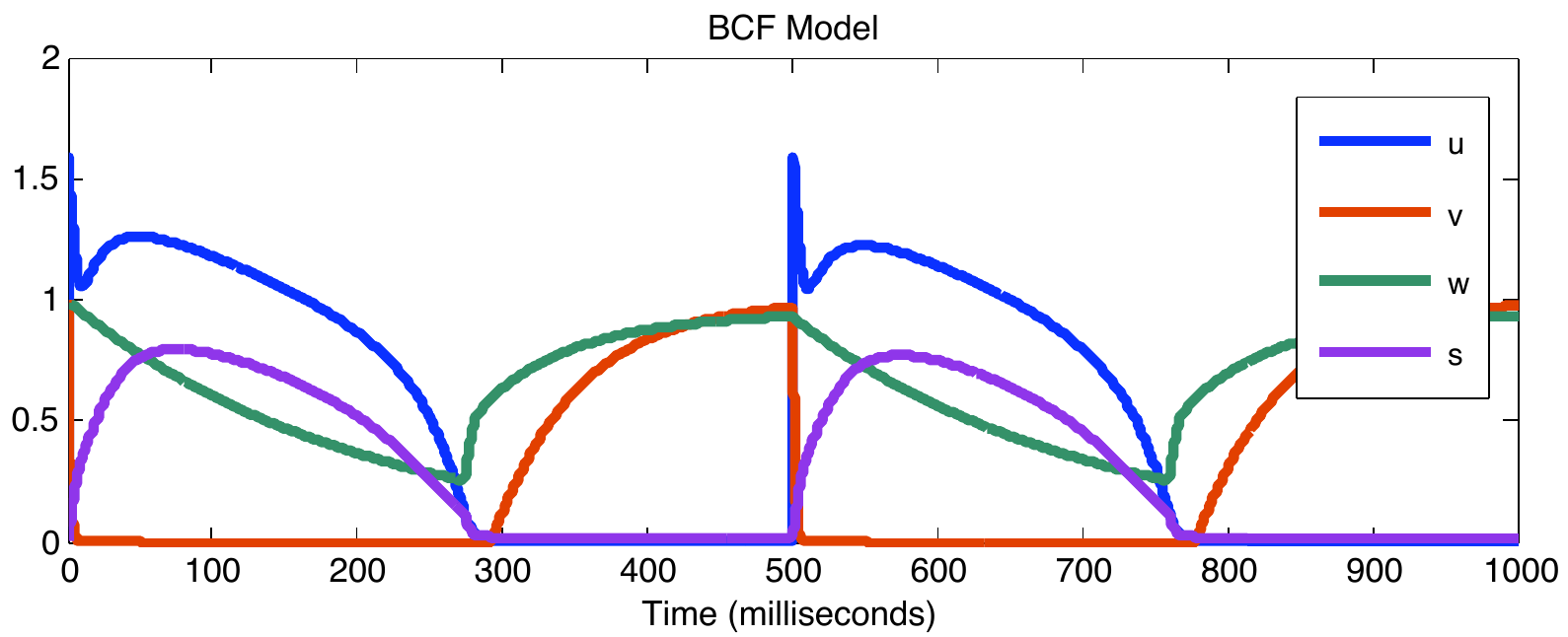}
}
\subfigure[]{
\includegraphics[width=8cm]{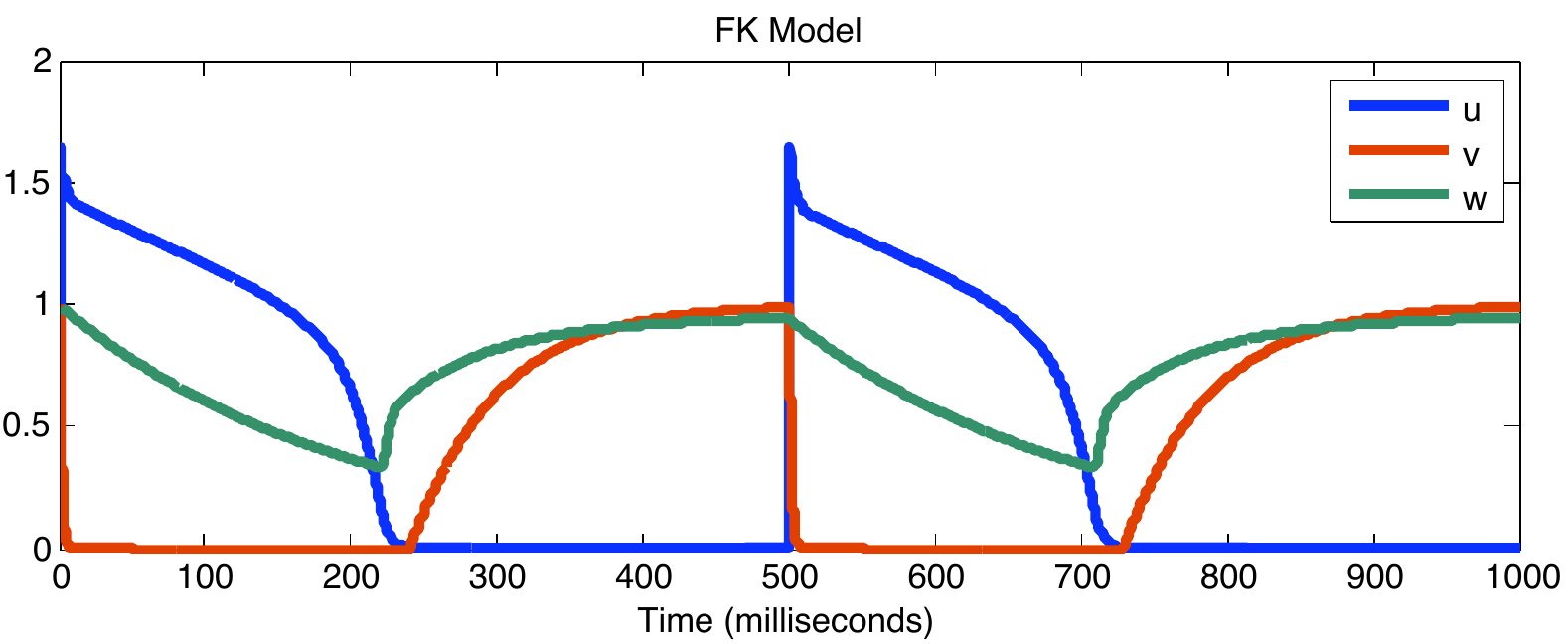}
}
\caption{The simulated witness trajectories of the BCF and the FK models.}
\label{trace}
\end{figure}

We specified both the BCF and the FK models using dReach's modeling language. 
Starting from the state ($u = 0$, $v = 1$, $w = 1$, $s = 0$, $\epsilon \in [0.9, 1.1]$) in Mode 1 (note that the value of $s$ does not matter to FK, which does not contains $s$), we checked whether Mode 4 is reachable using the parameter values presented in \cite{orovio08}. This was true (\ie, dReach returned $\delta$-$\mathsf{sat}$) for both models (Table \ref{tbl:exp}, Run\#1 and Run\#2).
The simulation of a few witness trajectories are shown in Figure \ref{trace} (the stimulus $\epsilon$ was reset every $500$ milliseconds).

\subsection{Model falsification}
\begin{figure}[th]
\centering
\includegraphics[scale=0.8]{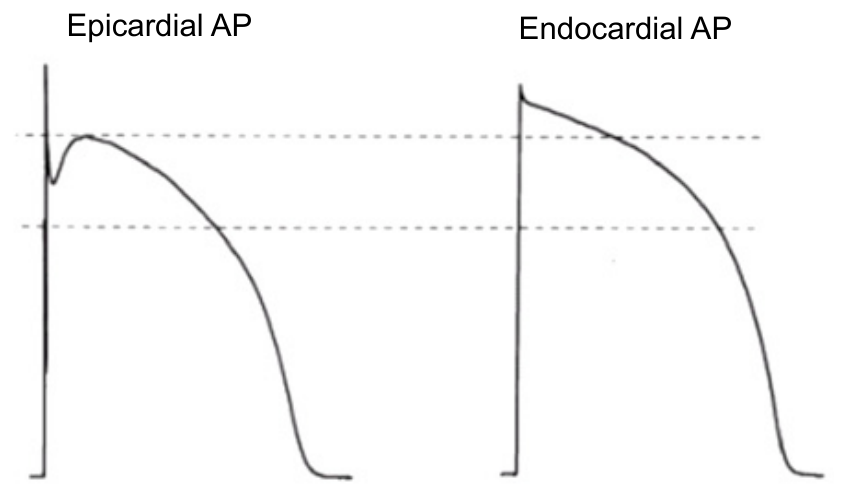}
\caption{Different AP morphologies observed in epicardial and endocardial cells \cite{nabauer96}.}
\label{ap}
\end{figure}

Both the BCF and the FK models were able to reproduce essential characteristics (\eg, steady-state action potential duration) observed in human ventricular cells \cite{fenton98,orovio08}. However, ventricular cells comprise three cell types, which possess different dynamical characteristics. For instance, Figure \ref{ap} shows that time courses of APs for epicardial and endocardial human ventricular cells have different morphologies \cite{nabauer96}. The important \textit{spike-and-dome} AP morphology can only be observed in epicardial cells but not endocardial cells. Hence, in a model-based study, one needs to identify cell-type-specific parameters to take account into cellular heterogeneity. The feasibility of this task will depend on the model of choice, as for certain models it would be impossible to reproduce a dynamical behavior no matter which parameter values are used. Here we illustrate that such models can be ruled out efficiently using our $\delta$-decision based parameter synthesis framework.

\paragraph{Robustness.}
We first considered the robustness property of the models. To ensure proper functioning of cardiac cells in noisy environments, an important property of the system is to filter out insignificant stimulation. Thus, we expected to see that AP could not be initiated for small $\epsilon$. Starting from the state ($u = 0$, $v = 1$, $w = 1$, $s = 0$, $\epsilon \in [0.0,0.25]$) in Mode 1, we checked the reachability of Mode 4. The $\mathsf{unsat}$ answer was returned by dReach for both the BCF and FK
model (Table \ref{tbl:exp}, Run\#3 and Run\#4), showing that the models are robust to stimulation amplitude.

\paragraph{AP morphology.}
Next we tested whether the models could reproduce the spike-and-dome AP morphology of epicardial cells. We introduced three auxiliary modes (Mode 5, 6 and 7). If $\text{\em time} \ge 1$, the system will jump from Mode 4 to Mode 5, in which $\epsilon$ will be reset to $0$. The system will jump from Mode 5 to Mode 6 if $\text{\em time} \ge 10$, and will jump from Mode 6 to Mode 7 if $\text{\em time} \ge 30$. In Modes 6 and 7, we enforced invariants $1.0 \le u \le 1.15$ and $1.18 \le u \le 2.0$, respectively, to depict the spike-and-dome morphology observed experimentally \cite{nabauer96}. We then checked reachability of Mode 7, starting from Mode 1 in state ($u = 0$, $v = 1$, $w = 1$, $s = 0$, $\epsilon \in [0.9,1.1]$, $\tau_{si} \in [1,2]$, $u_s \in [0.5,2]$),
where $\tau_{si}$ and $u_s$ are two model parameters that govern the dynamics of $u$ and $s$ in
Mode 3 and 4 (see Figure \ref{model}).
The $\delta$-$\mathsf{sat}$ answer
was returned for BCF  (Table \ref{tbl:exp}, Run\#5), while $\mathsf{unsat}$ was returned for FK (Table \ref{tbl:exp}, Run\#6), indicating that the FK model cannot reproduce spike-and-dome shapes using reasonable parameter values. Hence, FK is not suitable to study the dynamics of epicardial cells.

We remark that any $\mathsf{unsat}$ answer is guaranteed to be correct. This effectively
means that we proved that the FK model cannot reach Mode 7 for {\em any} starting state in the
rectangle ($u = 0$, $v = 1$, $w = 1$, $s = 0$, $\epsilon \in [0.9,1.1]$, $\tau_{si} \in [1,2]$,
$u_s \in [0.5,2]$). Sampling-based approaches cannot have the same level of certainty, while other
approaches cannot handle the complexity of the flows in the model.

\subsection{Parameter identification for cardiac disorders}

When the system cannot reach Mode 4, the cardiac cell loses excitability, which might lead to tachycardia or fibrillation. Starting with Mode 1, our task was to identify parameter ranges for which the system will never go into Mode 4. In what follows, we focused our study on the BCF model.
Grosu {\em et al.}~\cite{grosu11} have tackled this parameter identification problem by linearizing the BCF model into a piecewise-multiaffine system (referred as MHA). With this simplification, parameter ranges could be identified using the Rovergene tool \cite{rovergene}. However, the BCF and MHA models have different sets of parameters. Here we identify disease-related ranges of the {\em original} BCF parameters. It can be derived from the model equations that $\tau_{o1}$ and $\tau_{o2}$ govern the dynamics of $u$ in Mode 1 and Mode 2 respectively, and hence determine whether $\jump_{1\rightarrow 2}$ and  $\jump_{2\rightarrow 3}$ can be triggered. For $\tau_{o1}$, we performed a binary search in value domain $(0,0.01]$ to obtain a threshold value $\theta_{o1}$ such that Mode 4 is unreachable if $\tau_{o1} < \theta_{o1}$ while Mode 4 is reachable if $\tau_{o1} \ge \theta_{to1}$. The search procedure is illustrated in Algorithm \ref{bs}. Specifically, we set candidate $\theta^i_{o1}$ to be the midpoint of the search domain. We then checked the reachability of Mode 4 with the initial state ($u = 0$, $v = 1$, $w = 1$, $s = 0$, $\theta_{o1} = \theta^i_{o1}$). If $\delta$-$\mathsf{sat}$ was returned (\eg, Table \ref{tbl:exp}, Run\#7), we would recursively check the left-hand half of the search domain; otherwise (\eg, Table \ref{tbl:exp}, Run\#8), we would check the other half.

\begin{algorithm}
\SetAlFnt{\tiny}
\SetAlCapFnt{\small}
\SetAlCapNameFnt{\small}
\SetAlgoLined
\SetKwFunction{BinarySearch}{BinarySearch}
\SetKwFunction{dReach}{dReach}
\SetKwInOut{Input}{input}\SetKwInOut{Output}{output}
\BinarySearch{$M$, $v_{min}$, $v_{max}$, $\delta$}\\
\Input{A dReach model $M$; lower and upper bounds of parameter $v$: $v_{min}$, $v_{max}$; precision $\delta$}
\Output{A threshold value $\theta_{v}$}
\textbf{initialization}: $\theta_{v} \leftarrow (v_{min}+v_{max})/2$\;
\eIf{$|v_{min} - v_{max}| \le \delta$}{
  \Return $\theta_{v}$ \;
}{
  $Res \leftarrow $ \dReach{$M$, $\theta_{v}$, $\delta$} \;
  \eIf{$Res = \delta$-$\mathsf{sat}$}{
    \Return \BinarySearch{$M$, $v_{min}$, $\theta_{v}$, $\delta$}
  }{
    \Return \BinarySearch{$M$, $\theta_{v}$, $v_{max}$, $\delta$}
  }
}
\caption{Identify parameter threshold value using binary search. \label{bs}}
\end{algorithm}

\begin{figure}[h]
\centering
\includegraphics[scale=0.58]{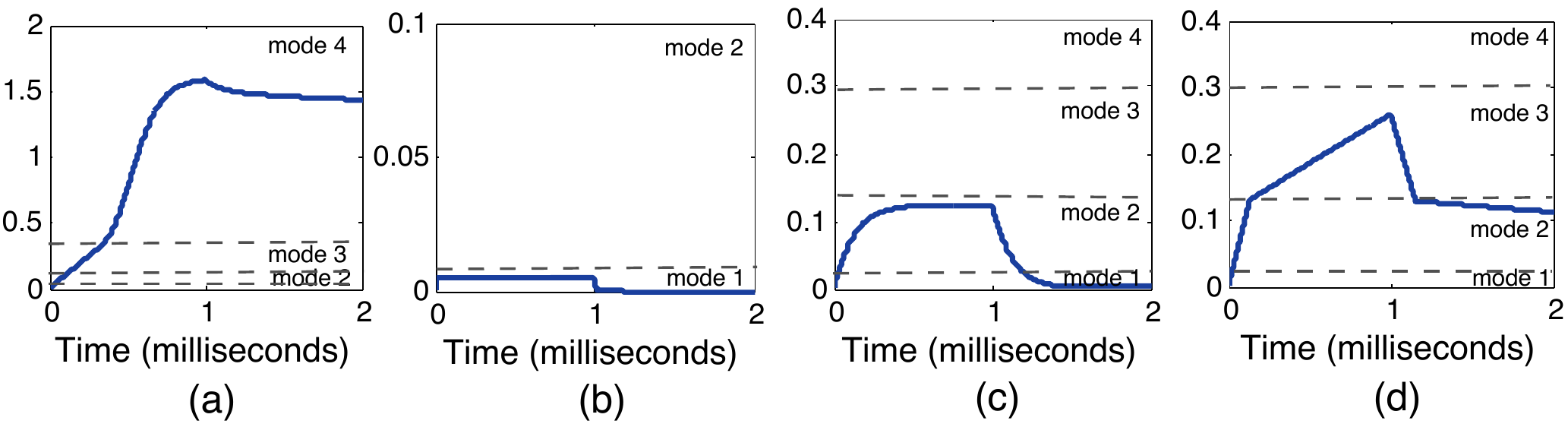}
\caption{Simulation results using disease related parameter values. (a) Normal condition (original parameters) (b) $\tau_{o1}=0.0055$ (c) $\tau_{o2} = 0.125$ (d) $\tau_{so1} =1.2$, $\tau_{so2} =1.0$ }
\label{cresults}
\end{figure}

In this manner, we identified $\theta_{o1}$ to be $0.006$, which suggest that when $\tau_{o1} \in (0, 0.006)$, the system will always stay in Mode 1 (Table \ref{tbl:exp}, Run\#9). Similarly, we also obtained a threshold value of $0.13$ for $\tau_{o2}$, such that Mode 3 cannot be reached when $\tau_{o2} \in (0, 0.13)$ (Table \ref{tbl:exp}, Run\#10). Furthermore, whether the system can jump from Mode 3 to Mode 4 depends on the interplay between $\tau_{so1}$ and $\tau_{so2}$.  For each value $\tau_{so2}^i$ of $\tau_{so2}$ sampled from domain $[0, 100]$, we performed the binary search in $[0, 5]$ to find the threshold value $\theta_{so1}$ such that Mode 4 is unreachable when $\tau_{so1} \in [0,\theta_{so1}]$ and $\tau_{so2} = {\tau_{so2}^i}$. By linear regression of the obtained values of $\theta_{so1}$, we identified one more condition that Mode 4 is unreachable:  $6.2 \cdot \tau_{so1} + \tau_{so2} \ge 9.9$ (\eg,  $\tau_{so1} \in [10, 40] \wedge \tau_{so1}\in [0.5, 2]$, see Table \ref{tbl:exp}, Run\#11). Taken together, we identified the following disease-related parameter ranges:
$$\tau_{o1} \in (0,0.006)\vee \tau_{o2} \in (0,0.13)\vee 6.2 \cdot \tau_{so1} + \tau_{so2} \ge 9.9$$
Figure \ref{cresults} visualizes these results by showing the simulated trajectories using corresponding parameter values.

{\small
\begin{table}[!th]
  \centering
  \small
  \begin{tabular}{c|c|c|c|c|c}
    \hline
    \hline
    Run & Model   & Initial State  & Var  & Result   & Time   \\
    \hline
    \hline
    1 & BCF & ($u = 0$, $v = 1$, $w = 1$, $s = 0$, $\epsilon \in [0.9,1.1]$) & 53  & $\delta$-$\mathsf{sat}$  & 303 \\
    2 & FK & ($u = 0$, $v = 1$, $w = 1$, $\epsilon \in [0.9,1.1]$)  & 53 & $\delta$-$\mathsf{sat}$ & 216 \\
    3 & BCF & ($u = 0$, $v = 1$, $w = 1$, $s = 0$, $\epsilon \in [0,0.25]$) & 53  & $\mathsf{unsat}$  & 2.09 \\
    4 & FK & ($u = 0$, $v = 1$, $w = 1$, $\epsilon \in [0.0,0.25]$)  & 53 & $\mathsf{unsat}$ & 0.78 \\
    5 & BCF & ($u = 0$, $v = 1$, $w = 1$, $s = 0$, $\epsilon \in [0.9,1.1]$)  & 89  & $\delta$-$\mathsf{sat}$  & 7,904 \\
    6 & FK & ($u = 0$, $v = 1$, $w = 1$, $\epsilon \mathord{\in} [0.9,1.1]$, $\tau_{si} \mathord{\in} [1,2]$, $u_{s} \mathord{\in} [0.5,2]$)  & 119 & $\mathsf{unsat}$ & 0.06 \\   
    7 & BCF & ($u = 0$, $v = 1$, $w = 1$, $s = 0$, $\tau_{o1} = 30.02$) & 53  & $\delta$-$\mathsf{sat}$  & 0.89 \\        
    8 & BCF & ($u = 0$, $v = 1$, $w = 1$, $s = 0$, $\tau_{o1} = 0.0055$) & 53  & $\mathsf{unsat}$  & 1.33 \\        
    9 & BCF & ($u = 0$, $v = 1$, $w = 1$, $s = 0$, $\tau_{o1} \in (0.0, 0.006)$) & 62  & $\mathsf{unsat}$  & 0.76 \\        
    10 & BCF & ($u = 0$, $v = 1$, $w = 1$, $s = 0$, $\tau_{o2} \in (0.0, 0.13)$)  & 62  & $\mathsf{unsat}$  & 0.32 \\     
    11 & BCF & ($u = 0$, $v = 1$, $w = 1$, $s = 0$, $\tau_{so1} \mathord{\in} [10, 40]$, $\tau_{so1}\mathord{\in} [0.5, 2]$) & 71  & $\mathsf{unsat}$  & 0.11 \\   
    \hline
    \hline
  \end{tabular}
  \caption{\small Experimental results.
    Var = number of variables in the unrolled formula,
    Result = bounded model checking result,
    Time = CPU time (s),
    $\delta=10^{-4}$.
}\label{tbl:exp}
\end{table}
}



\section{Conclusion}

We have presented a framework using $\delta$-complete decision procedures for the parameter identification 
of hybrid biological systems. We have used $\delta$-satisfiable formulas to describe parameterized hybrid automata 
and to encode parameter synthesis problems. We have employed $\delta$-decision procedures to perform bounded model 
checking, and developed an algorithm to obtain the resulting parameters. 
Our verified numerical integration and constraint programming algorithms effectively compute an over-approximation of the system dynamics. An  $\mathsf{unsat}$ answer can always be trusted, while a $\delta$-$\mathsf{sat}$ answer might be due to the over-approximation (see Section 2 for more details). We chose this behavior as it better fits with the safety requirements expected by formal verification.
We have demonstrated the applicability of our method on a highly nonlinear hybrid model of a cardiac cell that are
difficult to analyze with other verification tools. We have successfully ruled out a model candidate which did not fit experimental observations, and we have identified critical parameter ranges that can induce cardiac disorders.

It is worth noting that our method can be applied to ODE based models with discrete events, which are special forms of hybrid automata. Such models are often specified using the Systems Biology Markeup Language (SBML) 
and archived in the BioModels database \cite{biomodels}. Currently, we are currently developing an SBML-to-dReal translator to facilitate the $\delta$-decision based analysis of SBML models. 
Further, our method also has the potential to be applied to other model formalisms such as hybrid functional Petri nets \cite{hfpn} and the formalisms realized in Ptolemy \cite{ptolemy}. We plan to explore this in future work.
Another interesting direction is applying our method for parameter estimation from experimental data. By properly encoding the noisy wet-lab experimental data using logic formulas, bounded model checking can be utilized to find the unknown parameter values.
In this respect, the specification logic used in \cite{liu13} promises to offer helpful pointers.


\section*{Acknowledgements}
This work has been partially supported by award N00014-13-1-0090 of the US Office of Naval Research and award CNS0926181 of the National Science foundation (NSF).

\bibliographystyle{splncs}
\bibliography{bibliography}

\section*{Appendix: $\lrf$-Formulas and $\delta$-Decidability}

We will use a logical language over the real numbers that allows arbitrary {\em computable real functions}~\cite{CAbook}. We write $\lrf$ to represent this language. Intuitively, a real function is computable if it can be numerically simulated up to an arbitrary precision. For the purpose of this paper, it suffices to know that almost all the functions that are needed in describing hybrid systems are Type 2 computable, such as polynomials, exponentiation, logarithm, trigonometric functions, and solution functions of Lipschitz-continuous ordinary differential equations.

More formally, $\lrf = \langle \mathcal{F}, > \rangle$ represents the first-order signature over the reals with the set $\mathcal{F}$ of computable real functions, which contains all the functions mentioned above. Note that constants are included as 0-ary functions. $\lrf$-formulas are evaluated in the standard way over the structure $\mathbb{R}_{\mathcal{F}}= \langle \mathbb{R}, \mathcal{F}^{\mathbb{R}}, >^{\mathbb{R}}\rangle$. It is not hard to see that  we can put any $\lrf$-formula in a normal form, such that its atomic formulas are of the form $t(x_1,...,x_n)>0$ or $t(x_1,...,x_n)\geq 0$, with $t(x_1,...,x_n)$ composed of functions in $\mathcal{F}$. To avoid extra preprocessing of formulas, we can explicitly define $\mathcal{L}_{\mathcal{F}}$-formulas as follows.
\begin{definition}[$\lrf$-Formulas]
Let $\mathcal{F}$ be a collection of computable real functions. We define:
\begin{align*}
t& := x \; | \; f(t(\vec x)), \mbox{ where }f\in \mathcal{F} \mbox{ (constants are 0-ary functions)};\\
\varphi& := t(\vec x)> 0 \; | \; t(\vec x)\geq 0 \; | \; \varphi\wedge\varphi
\; | \; \varphi\vee\varphi \; | \; \exists x_i\varphi \; |\; \forall x_i\varphi.
\end{align*}
In this setting $\neg\varphi$ is regarded as an inductively defined operation
which replaces atomic formulas $t>0$ with $-t\geq 0$, atomic formulas $t\geq 0$
with $-t>0$, switches $\wedge$ and $\vee$, and switches $\forall$ and $\exists$.
\end{definition}
\begin{definition}[Bounded $\lrf$-Sentences]
We define the bounded quantifiers $\exists^{[u,v]}$ and $\forall^{[u,v]}$ as
$\exists^{[u,v]}x.\varphi =_{df}\exists x. ( u \leq x \land x \leq v \wedge
\varphi)$ and $
\forall^{[u,v]}x.\varphi =_{df} \forall x. ( (u \leq x \land x \leq v)
\rightarrow \varphi)$
where $u$ and $v$ denote $\lrf$ terms, whose variables only
contain free variables in $\varphi$ excluding $x$. A {\em bounded $\lrf$-sentence} is
$$Q_1^{[u_1,v_1]}x_1\cdots Q_n^{[u_n,v_n]}x_n\;\psi(x_1,...,x_n),$$
where $Q_i^{[u_i,v_i]}$ are bounded quantifiers, and $\psi(x_1,...,x_n)$ is
quantifier-free.
\end{definition}
\begin{definition}[$\delta$-Variants]\label{variants}
Let $\delta\in \mathbb{Q}^+\cup\{0\}$, and $\varphi$ an
$\lrf$-formula
$$\varphi: \ Q_1^{I_1}x_1\cdots Q_n^{I_n}x_n\;\psi[t_i(\vec x, \vec y)>0;
t_j(\vec x, \vec
y)\geq 0],$$ where $i\in\{1,...k\}$ and $j\in\{k+1,...,m\}$. The {\em
$\delta$-weakening} $\varphi^{\delta}$ of $\varphi$ is
defined as the result of replacing each atom $t_i > 0$ by $t_i >
-\delta$ and $t_j \geq 0$ by $t_j \geq -\delta$:
$$\varphi^{\delta}:\ Q_1^{I_1}x_1\cdots Q_n^{I_n}x_n\;\psi[t_i(\vec x, \vec
y)>-\delta; t_j(\vec x,
\vec y)\geq -\delta].$$
It is clear that $\varphi\rightarrow\varphi^{\delta}$~(see \cite{gao12b}).
\end{definition}
In~\cite{gao12a}, we have proved that the following $\delta$-decision problem is decidable, which is the basis of our framework.
\begin{theorem}[$\delta$-Decidability \cite{gao12a}]\label{delta-decide} Let $\delta\in\mathbb{Q}^+$ be
arbitrary. There is an algorithm which, given any bounded $\lrf$-sentence $\varphi$,
correctly returns one of the following two answers:
\begin{itemize}
\item $\delta$-$\mathsf{True}$: $\varphi^{\delta}$ is true.
\item $\mathsf{False}$: $\varphi$ is false.
\end{itemize}
When the two cases overlap, either answer is correct.
\end{theorem}
The following theorem states the (relative) complexity of the $\delta$-decision problem.
A bounded $\Sigma_n$ sentence is a bounded $\lrf$-sentence with $n$ alternating quantifier blocks 
starting with $\exists$. 
\begin{theorem}[Complexity \cite{gao12b}]\label{compmain}
Let $S$ be a class of $\lrf$-sentences, such that for any $\varphi$ in $S$, the terms in $\varphi$ are in Type 2 complexity class $\mathsf{C}$. Then, for any $\delta\in \mathbb{Q}^+$, the $\delta$-decision problem for bounded $\Sigma_n$-sentences in $S$ is in $\mathsf{(\Sigma_n^P)^C}$.
\end{theorem}
Basically, the theorem says that increasing the number of quantifier alternations will in general increase 
the complexity of the problem, unless $\mathsf{P}=\mathsf{NP}$ (recall that $\mathsf{\Sigma_0^P}=\mathsf{P}$ 
and $\mathsf{\Sigma_1^P}=\mathsf{NP}$).
This result can specialized for specific families of functions. For example, with polynomially-computable 
functions, the $\delta$-decision problem for bounded $\Sigma_n$-sentences is $\mathsf{(\Sigma_n^P)}$-complete.
For more details and results we again point the interested reader to \cite{gao12b}.

\end{document}